\documentclass[aps,prd,nofootinbib,onecolumn,a4paper]{revtex4-2}
\usepackage{bm,latexsym,amsmath,amssymb,amsfonts,mathrsfs,amsfonts}
\usepackage[dvipdfmx]{graphicx}
\usepackage[hang,small,bf]{caption}
\usepackage[subrefformat=parens]{subcaption}
\usepackage{ascmac}
\usepackage{physics}
\usepackage{color,float}
\usepackage{braket}
\usepackage{bbm}

\newcommand{\simgt}{\lower.5ex\hbox{$\; \buildrel > \over \sim \;$}}
\newcommand{\simlt}{\lower.5ex\hbox{$\; \buildrel < \over \sim \;$}}

\begin{document}
\title{Negativity volume of the generalized Wigner function in gravitating hybrid system}
\author{Daisuke Miki,$^{1}$ Akira Matsumura,$^{1}$ Kazuhiro Yamamoto$^{1,2,3}$}
\affiliation{$^1$Department of Physics, Kyushu University, 744 Motooka, Nishi-Ku, Fukuoka 819-0395, Japan}
\affiliation{
$^2$Research Center for Advanced Particle Physics, Kyushu University, 744 Motooka, Nishi-ku, Fukuoka 819-0395, Japan}
\email{miki.daisuke@phys.kyushu-u.ac.jp, \\
matsumura.akira@phys.kyushu-u.ac.jp, \\
yamamoto@phys.kyushu-u.ac.jp}
\affiliation{
$^3$International Center for Quantum-field Measurement
Systems for Studies of the Universe and Particles (QUP),
KEK, Oho 1-1, Tsukuba, Ibaraki 305-0801, Japan}
\date{\today}
\begin{abstract}
We evaluate the gravity-induced negativity volume of the generalized Wigner function in a hybrid system consisting of a particle in a two-localized superposition state and an oscillator.
The generalized Wigner function can capture
the nonclassicality of the system. 
The increase in the negativity volume of the generalized Wigner function can be an indicator of the entanglement generation, which is demonstrated in the hybrid system generating the gravity-induced entanglement in various initial states.
Moreover, by comparing the behaviors of the negativity volume with the entanglement fidelity, we show that the nonclassical feature of
entanglement is properly identified by the criterion based on the negativity volume of the generalized Wigner function when the oscillator is initially in a thermal state.
\end{abstract}
\maketitle
\section{INTRODUCTION}
Wigner function describes a distribution function in the phase space, which appears in various fields of sciences (e.g., \cite{Hu}).
It is known that the Wigner function defined for a particle in classical mechanics always takes positive values.
In contrast to classical mechanics, the Wigner function defined for a continuous variable system in quantum mechanics may have negative values, which is used as an indicator of the nonclassicality of the system \cite{Kenfact04}.
Such a property of the Wigner function 
can be measured in experiments, which is commonly used in quantum optics as a quantum signature 
\cite{Banaszek99,Lvovsky01,Zavatta04,Lvovsky09}.

The Wigner function for qubit systems has been extensively proposed
\cite{Tilma16,Rundle17,Tian18,Arkhipov18,Rundle19,Davies19}, which is defined on the Bloch sphere using the Euler angles. 
This Wigner function
enables us to analyze the discrete systems in the continuous phase space parameterized by the angles.
However, unlike the usual Wigner function for continuous variable systems, the Wigner function for qubit systems has negative regions on the Bloch sphere even for the arbitrary pure state of qubits.
Hence, the negative region of the Wigner function does not simply characterize the nonclassicality of the discrete system.
Thus, the Wigner function as an indicator of quantumness including qubit systems is not trivial. 
Under such circumstances, Ref. \cite{Arkhipov18} introduced the negativity volume of the generalized Wigner function as an indicator of the nonclassical feature for a hybrid system.

The hybrid system, consisting of the discrete and continuous variable subsystems, is one of the central research in quantum information science; quantum information processing \cite{Andersen15}, quantum teleportation \cite{Gisin07,Ulanov17}, and quantum entanglement \cite{Kreis12,Jeong14}.
The generalized Wigner function in the hybrid system is obtained in Ref. \cite{Tilma16}, but it is demonstrated that the negative values do not necessarily characterize the nonclassicality because of the property of the Wigner function in qubit systems.
However, Ref. \cite{Arkhipov18} showed that there is a critical value for the negativity volume of the Wigner function for the separable hybrid system; if the negativity volume is larger than the critical value, nonclassicality appears.
We note that this criterion is a sufficient condition, not a necessary condition.

In the present study, we consider
the generalized Wigner function of the hybrid system consisting of a particle in a two-state superposition and an oscillator, which are gravitationally interacting. 
Refs.~\cite{Carney21,Strltsov21,Ma21,MY21} discussed the quantum entanglement through the gravitational interaction in this hybrid system.
Recently, the experimental proposal of the gravity-induced entanglement between two particles has attracted interest \cite{Bose,Marletto}.
The quantum entanglement is known as a nonlocal quantum correlation, which is not newly generated by classical processes (e.g. \cite{Horodecki09}).
Refs.~\cite{Bose,Marletto} showed that two particles in a superposition state are entangled if Newtonian gravity obeys quantum mechanics, which was shown as a feasible experimental proposal in the future.
This two-particle model has been the subject of several advanced studies \cite{Nguyen20,Miki20,Matsumura21,Sugiyama21,Feng22}.
In addition, gravitational entanglement between mechanical oscillators in optomechanical systems has also been investigated \cite{Miao20,Krisnanda20,Matsumura20,Datta21,Miki21,Plato22};
experiments \cite{Matsumoto19,Catano20,Matsumoto20} and theories  \cite{Chen13,Bowen15,Rudolph22,Miki22,Miki23} have been developed towards the realization of macroscopic quantum states to verify the quantum nature of gravity.
Furthermore, quantum entanglement generation by Newtonian gravity has been discussed from the viewpoint of quantum field theory and gravitons \cite{Belenchia18,Marshman20,Carney22,Danielson22,Bose22,Hidaka22,Sugiyama22,Biswas22,Iso22,Sugiyama23}.

For the hybrid system considered in the present paper, we evaluate the gravity-induced negativity volume of the generalized Wigner function as a quantity to characterize the quantumness of gravity and discuss its correspondence with 
the condition for the quantum entanglement  
based on the generalized Wigner function \cite{Arkhipov18}.
We discuss the applicability of the quantum entanglement criterion with the generalized Wigner function in comparison with the entanglement fidelity criterion proposed in Ref.~\cite{Roszak15,Roszak18,Strzalka20}.
In particular, we show that the condition based on the generalized Wigner function correctly captures the entanglement even when the oscillator is initially in a thermal state.

The present paper is organized as follows:
In Sec.~II, we briefly review the entanglement criterion
with the generalized Wigner function for a hybrid system,
which was proposed in Ref.~\cite{Arkhipov18}. 
In Sec.~III, after briefly reviewing the hybrid system 
to detect the gravity-induced entanglement considered in 
Refs.~\cite{Carney21,MY21}, we evaluate the evolution of the negativity volume in the hybrid system for three different initial states to determine the entanglement generation. 
In Sec.~IV, we evaluate the entanglement fidelity for the three 
states introduced in Sec.~III, to verify that the criterion with the negativity volume works properly. 
Sec.~V is devoted to summary and conclusions. 
In Appendix A, we introduce the generalized Wigner function
and explain the negativity volume for a hybrid system following Ref.~\cite{Arkhipov18}.
\section{WIGNER FUNCTION FOR HYBRID SYSTEM}
In this section, we introduce the generalized Wigner function for the hybrid system described by a continuous variable and a discrete variable.
We first consider the Wigner function for the continuous variable system defined by
\begin{align}
	\label{wig:cv}
	W_{b}
	&=\text{Tr}[\rho_{b}\hat{\Delta}_{b}],\quad
	\hat{\Delta}_{b}
	=\hat{D}_{b}\hat{\Pi}_{b}\hat{D}_{b}^{\dagger},
\end{align}
where $\rho_{b}$ is the density matrix, $\hat{D}_{b}=e^{\beta \hat{b}^{\dagger}-\beta^*\hat{b}}$ is the displacement operator, and $\hat{\Pi}_{b}=\frac{2}{\pi}e^{i\pi\hat{b}^{\dagger}\hat{b}}$ is the parity operator, and
$\hat{b}$ and $\hat{b}^{\dagger}$ are the creation and annihilation operators.
It is well known that this Wigner function characterizes the nonclassical nature of continuous variable systems and that it has negative values in quantum states such as a Shr\"{o}dinger-cat state.

On the other hand, the Wigner function for the discrete system is defined to satisfy the Stratonovich-Weyl conditions as follows \cite{Tilma16,Rundle17,Tian18,Arkhipov18,Rundle19,Davies19}:
\begin{align}
	\label{wig:dv}
	W_{q}
	&=\text{Tr}[\rho_{q}\hat{\Delta}_{q}],\quad
	\hat{\Delta}_{q}
	=\hat{U}_{q}\hat{\Pi}_{q}\hat{U}_{q}^{\dagger},
\end{align}
where $\hat{U}_{q}=e^{\frac{i}{2}\phi\hat{\sigma}_{z}}e^{\frac{i}{2}\theta\hat{\sigma}_{y}}e^{\frac{i}{2}\Phi\hat{\sigma}_{z}}$ is the unitary operator of SU(2) and $\hat{\Pi}_{q}=\frac{1}{2}(\hat{\mathbbm{1}}-\sqrt{3}\hat{\sigma}_{z})$ is the parity operator.
$\phi,\Phi\in[0,2\pi],\theta\in[0,\pi]$ are the Euler angles and $\hat{\sigma}_{y},~\hat{\sigma}_{z}$ are the Pauli operators.
The negative value of the Wigner function of the discrete system does not simply characterize the quantum nature of the system.
For the general pure qubit state,
\begin{align}
\label{wig:pq}
\rho_{q}^{p}
&=\ket{q}\bra{q},\quad
\ket{q}=\sqrt{\alpha}\ket{0}+e^{i\chi}\sqrt{1-\alpha}\ket{1},\quad
\alpha\in[0,1],\quad\chi\in[0,2\pi],
\end{align}
we derive the Wigner function as
\begin{align}
\label{wig:pqwig}
W_{q}[\rho_{q}^{p}]
&=\sqrt{3\alpha(1-\alpha)}\sin\theta\cos(\chi+2\phi)+\sqrt{3}(1-2\alpha)/2 \cos\theta+1/2.
\end{align}
Considering the state $\ket{q}=\ket{0}$ for $\alpha=1$, the Wigner function reduces to
\begin{align}
	\label{wig:dv0}
	W_{q}[\ket{0}\bra{0}]
	&=\frac{1}{2}
	\left(
	1-\sqrt{3}\cos\theta
	\right),
\end{align}
which has the negative values on the Bloch sphere.
Hence, the Wigner function of the discrete system has negative values even for a pure state.

To characterize the nonclassicality of the hybrid system,
the authors of Ref. \cite{Arkhipov18} introduced the negativity volume of the generalized Wigner function as
\begin{align}
	\label{wig:nv}
	\mathcal{V}[\rho]
	&=\frac{1}{2}
	\left(
	\int d\phi d\theta d^{2}\beta\frac{\sin\theta}{2\pi}|W|-1
	\right),
\end{align}
where $W$ is the generalized Wigner function defined by
\begin{align}
	\label{wig:hy}
	W[\rho]
	=&\text{Tr}[\rho\hat{\Delta}_{q}\otimes\hat{\Delta}_{b}].
\end{align}
The Wigner function and the negativity volume for the reduced density matrix for the continuous variable system and for the qubit are respectively defined as
\begin{align}
	\label{wig:nvb}
	W_{b}
	&=\text{Tr}[\rho_{b}\Delta_{b}],
	\qquad
	\mathcal{V}_{b}[\rho_{b}]=\frac{1}{2}
	\left(
	\int d^{2}\beta |W_{b}|-1
	\right),\\
	\label{wig:nvq}
	W_{q}
	&=\text{Tr}[\rho_{q}\Delta_{q}],
	\qquad
	\mathcal{V}_{q}[\rho_{q}]=\frac{1}{2}
	\left(
	\int \frac{\sin\theta}{2\pi}d\phi d\theta |W_{q}|-1
	\right),
\end{align}
where $\rho_{b}=\text{Tr}_{q}[\rho]$ and $\rho_{q}=\text{Tr}_{b}[\rho]$. The authors of 
Ref.~\cite{Arkhipov18} showed that the negativity volume of the arbitrary pure qubit state \eqref{wig:pq} is the same value
\begin{align}
	\label{wig:nvqp}
	\mathcal{V}_{q}[\rho_{q}^{p}]
	&=\frac{1}{\sqrt{3}}-\frac{1}{2},
\end{align}
and suggested that this negativity volume is the upper bound for any qubit state.
They also expected that the entanglement criterion for the hybrid system is expressed as follows; the state is entangled if the inequality
\begin{align}
	\label{wig:criterion}
	\mathcal{V}[\rho]
	&\le
	\mathcal{V}_{\text{cr}}
	=\frac{2}{\sqrt{3}}\sum_{i}p_{i}\mathcal{V}_{b}[\rho_{b}^{i}]+\frac{1}{\sqrt{3}}-\frac{1}{2},
\end{align}
is violated.
Here, the critical value $\mathcal{V}_{\text{cr}}$ corresponds to the maximum value for the separable hybrid system written as $\rho^{\text{sep}}=\sum_{i}p_{i}\rho_{q}^{i}\otimes\rho_{b}^{i}$.
Hence, this criterion indicates that when the negativity volume is larger than the critical value, additional nonclassicality appears, namely entanglement.

On the other hand, there is no upper limit on the negativity volume for the reduced continuous variable system $\mathcal{V}_{b}$.
There is the possibility in which a basis transformation can result in different values of the reduced negativity volume, even for the same state.
For example, we consider the state 
$\ket{\phi}\propto\ket{0}_q\ket{\alpha}_b+\ket{1}_q\ket{-\alpha}_b$, where $\ket{0}_q$ and $\ket{1}_q$
are the states of a qubit, and $\ket{\pm\alpha}_b$ is the coherent state of a continuous variable. 
Tracing over the qubit system, the reduced state is mixed state $\rho_{b}\propto\ket{\alpha}{\hspace{-1mm}}{}_b{}_b{\hspace{-1mm}}\bra{\alpha}+\ket{-\alpha}{\hspace{-1mm}}{}_b{}_b{\hspace{-1mm}}\bra{-\alpha}$ and the reduced negativity volumes $\mathcal{V}_{b}[\ket{\alpha}{\hspace{-1mm}}{}_b{}_b{\hspace{-1mm}}\bra{\alpha}]$ and $\mathcal{V}_{b}[\ket{-\alpha}{\hspace{-1mm}}{}_b{}_b{\hspace{-1mm}}\bra{-\alpha}]$ are zero.
However, we consider the transformation of the basis as $\ket{+}_q=\ket{0}_q+\ket{1}_q$ and 
$\ket{-}_q=\ket{0}_q-\ket{1}_q$, and the reduced state is written in the mixed state $\rho_{b}\propto(\ket{\alpha}{\hspace{-1mm}}{}_b+\ket{-\alpha}{\hspace{-1mm}}{}_b)({}_b{\hspace{-1mm}}\bra{\alpha}+{}_b{\hspace{-1mm}}\bra{-\alpha})+(\ket{\alpha}{\hspace{-1mm}}{}_b-\ket{-\alpha}{\hspace{-1mm}}{}_b)({}_b{\hspace{-1mm}}\bra{\alpha}-{}_b{\hspace{-1mm}}\bra{-\alpha})$.
Then, the reduced negativity volumes $\mathcal{V}_{b}[(\ket{\alpha}{\hspace{-1mm}}{}_b+\ket{-\alpha}{\hspace{-1mm}}{}_b)({}_b{\hspace{-1mm}}\bra{\alpha}+{}_b{\hspace{-1mm}}\bra{-\alpha})]$ and $\mathcal{V}_{b}[(\ket{\alpha}{\hspace{-1mm}}{}_b-\ket{-\alpha}{\hspace{-1mm}}{}_b)({}_b{\hspace{-1mm}}\bra{\alpha}-{}_b{\hspace{-1mm}}\bra{-\alpha})]$ are non-zero.
Thus, the treatment of negativity volume for the reduced
system of a continuous variable system coupled to a qubit is non-trivial.
Despite such a problem, in the present paper, 
we demonstrate that the criterion based on the negativity volume of the generalized Wigner function is efficient by comparing the fidelity-based condition for various initial
states such as superposed qubit-coherent, -thermal, and -cat state.
We omit the subscript $"b"$ and $"q"$ for simplicity.

\section{negativity volume AND ENTANGLEMENT INDUCED BY GRAVITY IN HYBRID SYSTEM}
\begin{figure}[t]
	\centering
	\includegraphics[width=14.cm]{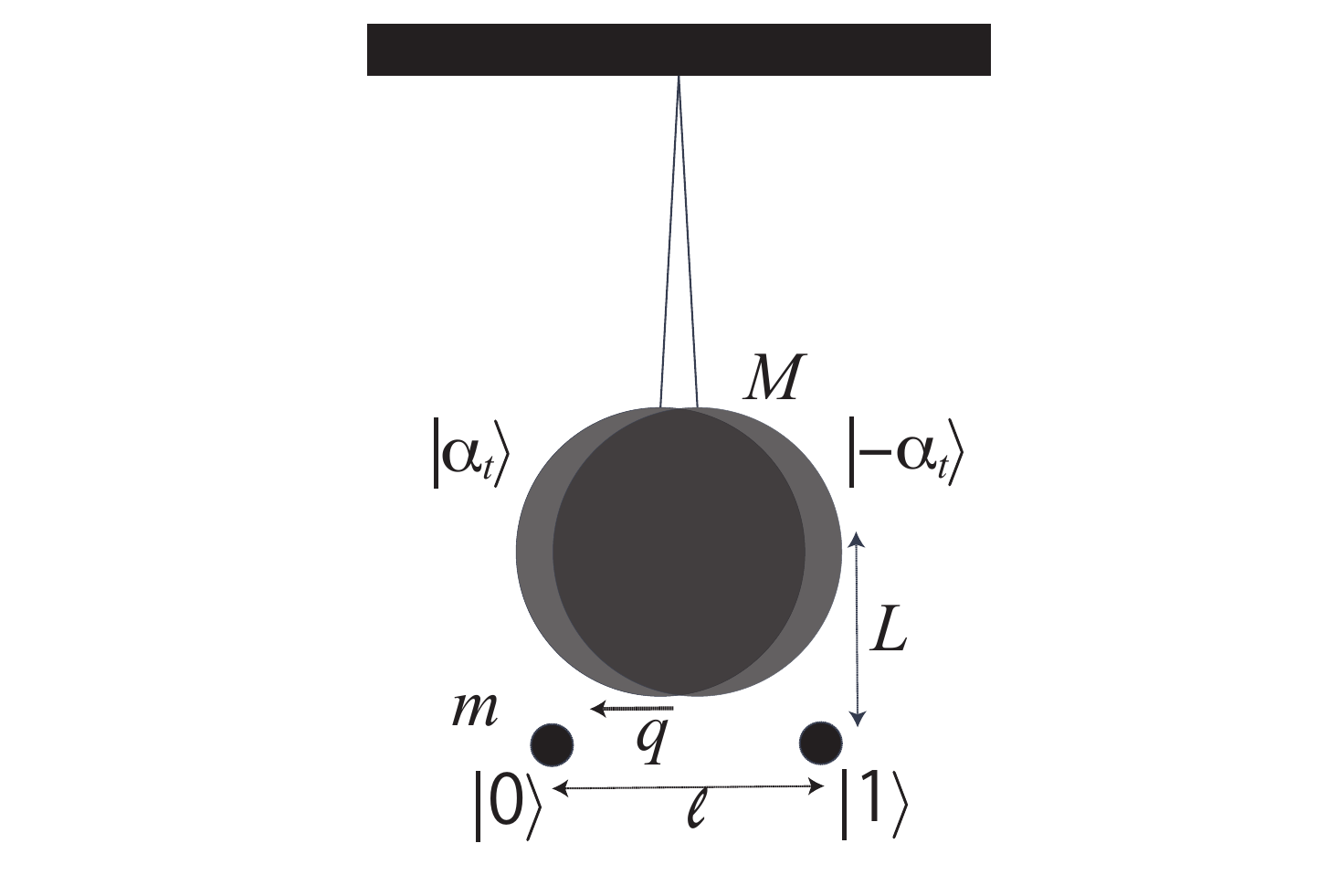}
	\caption{
   Schematic plot of the model as a hybrid system. 
   This figure assumes $\gamma=0$ in Eq.~(\ref{q-coh:phi}).}
    \label{fig:schematic}
\end{figure}

In this section, we consider the hybrid system consisting of a spin particle with mass $m$ and an oscillator with mass $M$ considered in Refs.~\cite{Carney21,MY21}. Figure \ref{fig:schematic} shows a schematic plot of the system.
The spin particle is prepared as the superposition state of the two spatially localized states separated by distance $\ell$ by using the Stern-Gerlach scheme. 
Then, the oscillator also becomes the superposition state due to gravitational interaction with the spatially-localized 
superposed particle.
The Hamiltonian of the hybrid system is
\begin{align}
	\label{hy:Hamiltonian}
	\hat{H}
	&=\omega\hat{a}^{\dagger}\hat{a}+\hat{H}_{g},
\end{align}
where $\hat{a},~\hat{a}^{\dagger}$ are the creation and annihilation operators and $\omega$ is the frequency of the oscillator.
Here, we ignore the Larmor precession for simplicity.
Hamiltonian $\hat{H}_{g}$ describes the gravitational interaction given by
\begin{align}
	\label{hy:Hg}
	H_{g}
	&=-\frac{GMm}{\sqrt{L^{2}+(\hat{q}'+\hat{\sigma}_{z}\ell/2)^2}}
	\simeq
	\frac{GMm\ell}{\sqrt{L^{2}+\ell^{2}/4}^{3}}\hat{\sigma}_{z}\hat{q}'+\text{const.},
\end{align}
where $\hat{q}$ is the position operator of the oscillator, $L$ is the distance between the spin particle and the oscillator, $\ell$ is the distance between the positions of the superposed particle and $G$ is the gravitational constant.
Here, we define the dimensionless position operator $\hat{q}=\sqrt{M\omega}\hat{q}'$ and introduce the gravitational coupling $g$ as
\begin{align}
	\label{hy:gconst}
	g
	&=\frac{GMml}{\sqrt{L^{2}+\ell^{2}/4}^{3}}\frac{1}{\sqrt{2M\omega}}.
\end{align}
Hamiltonian $H_{g}$ is rewritten as $\hat{H}_{g}=\sqrt{2}g\hat{\sigma}_{z}\hat{q}'$ and the unitary evolution is
\begin{align}
	\label{hy:unitary}
	\hat{U}(t)
	&=e^{-i\hat{H}t}
	=e^{-i\omega\hat{a}\hat{a}^{\dagger}t}e^{-\hat{\sigma}_{z}(\alpha^{*}_{t}\hat{a}^{\dagger}-\alpha_{t}\hat{a})+iC^{2}_{t}},
\end{align}
where we defined
\begin{align}
	\label{hy:ab}
	\alpha_{t}
	&=\lambda(e^{-i\omega t}-1),\quad
	C_{t}
	=\lambda^{2}(\omega t-\sin[\omega t]),\quad
	\text{with}\quad
	\lambda=\frac{g}{\omega}.
\end{align}
	
\subsection{INITIAL QUBIT-COHENRENT STATE}
\begin{figure}[t]
	\centering
	\includegraphics[width=10.5cm]{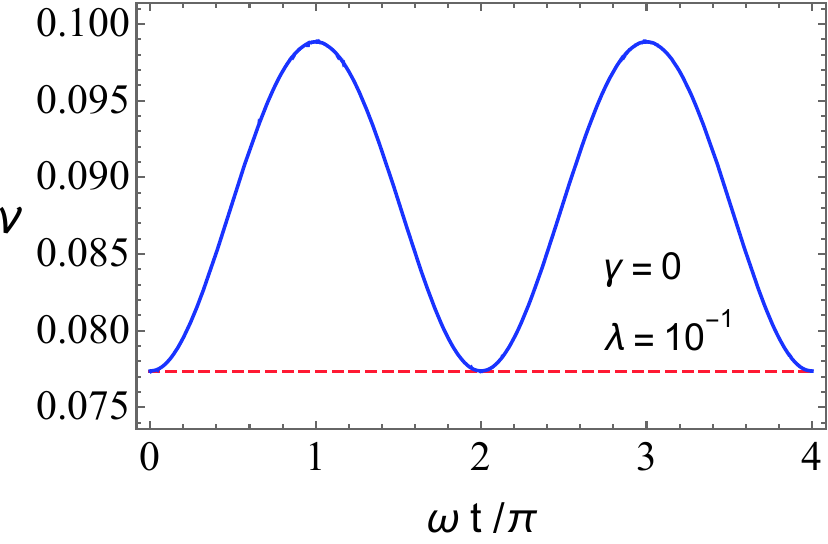}
	\caption{
	The behavior of the negativity volume as a function of dimensionless time $\omega t/\pi$ for $\lambda=10^{-1}, \gamma=0$.
	The blue solid curve and the red dashed line represent the negativity volume $\mathcal{V}[\rho(t)]$ and the critical value $\mathcal{V}_{cr}=1/\sqrt{3}-1/2$, respectively.
	}
	\label{fig:q-coh}
\end{figure}

We assume that the particle and the oscillator are initially in a superposed state and a coherent state, respectively.
The state of the total system is
\begin{align}
	\label{q-coh:ini}
	\ket{\phi}
	&=\frac{1}{\sqrt{2}}(\ket{0}+\ket{1})\otimes\ket{\gamma},
\end{align}
which evolves by the Hamiltonian \eqref{hy:Hamiltonian} as
\begin{align}
	\label{q-coh:phi}
	\ket{\phi(t)}
	&=\frac{1}{\sqrt{2}}
	\left(
	e^{-(\alpha^{*}_{t}\gamma^{*}-\alpha_{t}\gamma)/2}\ket{0}\ket{\alpha_{t}+\gamma_{t}}
	+e^{(\alpha^{*}_{t}\gamma^{*}-\alpha_{t}\gamma)/2}\ket{1}\ket{-\alpha_{t}+\gamma_{t}}
	\right),
\end{align}
where $\gamma_{t}=e^{-i\omega t}\gamma$ and we ignore the global phase.
The generalized Wigner function of the state $\rho(t)=\ket{\phi(t)}\bra{\phi(t)}$ is given by
\begin{align}
	\label{q-coh:W}
	W[\rho(t)]
	&=\frac{1}{2\pi}\left((1-\sqrt{3}\cos\theta)e^{-2|\beta-\alpha_{t}-\gamma_{t}|^{2}}+(1+\sqrt{3}\cos\theta)e^{-2|\beta+\alpha_{t}-\gamma_{t}|^{2}}
	\right.\notag\\&\left.\qquad
	+2\sqrt{3}\sin\theta e^{-2|\beta-\gamma_{t}|^{2}}\cos(\phi+4\text{Im}[\alpha_{t}^{*}\beta])\right).
\end{align}
Here, the reduced density matrix of the oscillator $\rho_{b}(t)=\text{Tr}_{q}[\rho(t)]$ is
\begin{align}
	\label{q-coh:red}
	\rho_{b}(t)
	&=\frac{1}{2}(\rho_{b1}(t)+\rho_{b2}(t)),\\
	\rho_{b1}(t)
	&=\ket{\alpha_{t}+\gamma_{t}}\bra{\alpha_{t}+\gamma_{t}},\quad
	\rho_{b2}(t)
	=\ket{-\alpha_{t}+\gamma_{t}}\bra{-\alpha_{t}+\gamma_{t}},
\end{align}
and the corresponding Wigner functions are\begin{align}
	\label{q-coh:Wred}
	W_{b}[\rho_{b1}(t)]
	&=\frac{1}{2\pi}e^{-2|\beta-\alpha_{t}-\gamma_{t}|^{2}},\quad
	W_{b}[\rho_{b2}(t)]
	=\frac{1}{2\pi}e^{-2|\beta+\alpha_{t}-\gamma_{t}|^{2}}.
\end{align}
Because this state is the mixed state of two Gaussian states, the negativity volume of the reduced oscillator's state is zero.
Hence, if the inequality
\begin{align}
	\label{q-coh:ineq}
	\mathcal{V}[\rho(t)]
	&\le\frac{1}{\sqrt{3}}-\frac{1}{2},
\end{align}
is violated, the state is 
expected to be entangled.
Fig. \ref{fig:q-coh} shows the behavior of the negativity volume for $\gamma=0$.
We here use $\lambda=10^{-1}$ to easily understand the behavior of the negativity volume of the Wigner function, although $\lambda$ is much smaller in realistic situations.
The red dashed line represents the critical value $\mathcal{V}_{\text{cr}}$ in the right-hand side of the inequality \eqref{q-coh:ineq}.
The negativity volume at the initial time in the separable state is
the same as the critical value.
The result is consistent with the criterion that the state is entangled when the negativity volume is larger than the critical value $1/\sqrt{3}-1/2$
in this case.
	
\subsection{INITIAL QUBIT-THERMAL STATE}
We next consider the thermal state as the initial state of the oscillator.
In a realistic situation assuming an experiment, 
the thermal effect should be taken into account, and the generation of the gravity-induced entanglement is non-trivial. The entanglement in the hybrid system under the thermal effect has not been explicitly estimated \cite{Carney21,MY21}. 
We perform the estimation through the negativity volume of the generalized Wigner function of the hybrid system. 

\begin{figure}[t]
	\centering
	\includegraphics[width=10.5cm]{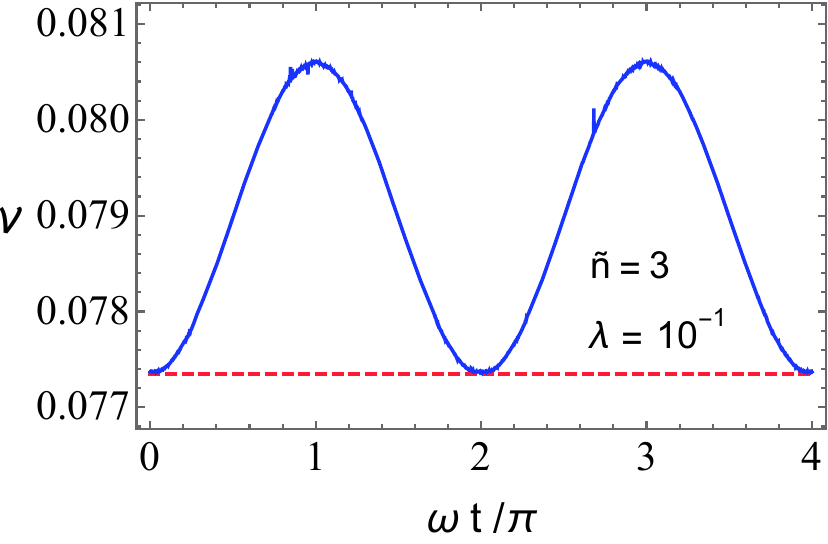}
	\caption{
   The behavior of the negativity volume (blue curve) as a function of dimensionless time $\omega t/\pi$ for the initial qubit-thermal state with $\lambda=10^{-1},\bar{n}=3$.
	The negativity volume exhibits the same periodic behavior as that in the initial qubit-coherent state of Fig.~\ref{fig:q-coh}, though the peak value is smaller.
 The red dashed line is the critical value $1/\sqrt{3}-1/2$
in this case.}
	\label{fig:q-th}
\end{figure}

The initial density matrix of the total system is
\begin{align}
	\label{q-th:ini}
	\rho^{th}(0)
	&=\frac{1}{2}(\ket{0}+\ket{1})(\bra{0}+\bra{1})\otimes\frac{1}{\pi\bar{n}}\int d^{2}\gamma e^{-|\gamma|^{2}/\bar{n}}\ket{\gamma}\bra{\gamma},
\end{align}
where $\bar{n}=k_{B}T/2\omega$ with the temperature $T$ is thermal phonon occupancy number.
The evolution of the density matrix by the Hamiltonian \eqref{hy:Hamiltonian} is
\begin{align}
	\label{q-th:rho}
	\rho^{th}(t)
	&=\frac{1}{2\pi\bar{n}}\int d^{2}\gamma e^{-|\gamma|^{2}/\bar{n}}
	\left\{
	\ket{0}\ket{\alpha_{t}+\gamma_{t}}\bra{0}\bra{\alpha_{t}+\gamma_{t}}
	+e^{-(\alpha_{t}^{*}\gamma^{*}-\alpha_{t}\gamma)}\ket{0}\ket{\alpha_{t}+\gamma_{t}}\bra{1}\bra{-\alpha_{t}+\gamma_{t}}\right.\notag\\
	&\qquad\qquad\qquad\left.
	+e^{\alpha_{t}^{*}\gamma^{*}-\alpha_{t}\gamma}\ket{1}\ket{-\alpha_{t}+\gamma_{t}}\bra{0}\bra{\alpha_{t}+\gamma_{t}}
	+\ket{1}\ket{-\alpha_{t}+\gamma_{t}}\bra{1}\bra{-\alpha_{t}+\gamma_{t}}
	\right\},
\end{align}
and the generalized Wigner function is
\begin{align}
	\label{q-th:W}
	W[\rho^{th}(t)]
	=\frac{1}{2\pi(2\bar{n}+1)}
	&\left\{
	(1-\sqrt{3}\cos\theta)e^{-\frac{2}{2\bar{n}+1}|\beta-\alpha_{t}|^{2}}
	+(1+\sqrt{3}\cos\theta)e^{-\frac{2}{2\bar{n}+1}|\beta+\alpha_{t}|^{2}}\right.\notag\\
	&\left.+2\sqrt{3}e^{-\frac{2}{2\bar{n}+1}|\beta|^{2}}\sin\theta\cos[\phi+4\text{Im}[\alpha_{t}^{*}\beta]]
	\right\}.
\end{align}
Then, the reduced density matrix of the oscillator $\rho^{th}_{b}$ is
\begin{align}
	\rho^{th}_{b}(t)
	&=\frac{1}{2}(\rho^{th}_{b1}(t)+\rho^{th}_{b2}(t)),\notag\\
	\rho^{th}_{b1}(t)
	&=\frac{1}{\pi\bar{n}}\int d^{2}\gamma e^{-|\gamma|^{2}/\bar{n}}
	\ket{\alpha_{t}+\gamma_{t}}\bra{\alpha_{t}+\gamma_{t}},\quad
	\rho^{th}_{b2}(t)
	=\frac{1}{\pi\bar{n}}\int d^{2}\gamma e^{-|\gamma|^{2}/\bar{n}}
	\ket{-\alpha_{t}+\gamma_{t}}\bra{-\alpha_{t}+\gamma_{t}},\notag
\end{align}
and the Wigner function is obtained by
\begin{align}
	\label{Wth:reduced}
	W_{b}[\rho_{b1}^{th}]
	&=\frac{1}{2\pi(2\bar{n}+1)}e^{-\frac{1}{2\bar{n}+1}|\beta-\alpha_{t}|^{2}},\quad
	W_{b}[\rho_{b2}^{th}]
	=\frac{1}{2\pi(2\bar{n}+1)}e^{-\frac{1}{2\bar{n}+1}|\beta+\alpha_{t}|^{2}}.
\end{align}
These Wigner functions of the reduced state are always non-negative and the negativity volume is zero, $\mathcal{V}[\rho_{b1}^{th}]=\mathcal{V}[\rho_{b2}^{th}]=0$.
Then, the inequality \eqref{wig:criterion} reduces to the inequality \eqref{q-coh:ineq}.
Hence, the violation of the inequality \eqref{q-coh:ineq} 
is the criterion for the entanglement generation due to gravity.
Figure \ref{fig:q-th} shows the behavior of the negativity volume and critical value for $\lambda=10^{-1},\bar{n}=3$.
The entanglement behavior for the initial qubit-thermal state is similar to that for the initial qubit-coherent state, but the maximum value of the negativity volume is smaller due to the thermal effects.
However, the criterion based on the negativity volume of the generalized Wigner function detects the entanglement even for the mixed state in the hybrid system.

\begin{figure}[tbp]
	\centering
	\includegraphics[width=10.5cm]{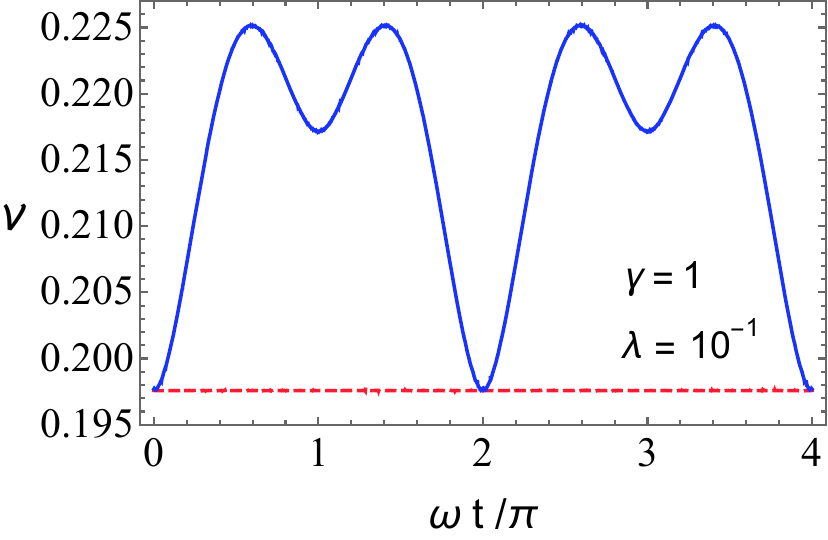}
	\caption{
	The negativity volume for the initial qubit-cat state with $\lambda=10^{-1},\gamma=1$. The red dashed curve is the right-hand side of Eq.~(\ref{q-cat:criterion}).
	}
	\label{fig:q-cat}
\end{figure}

\subsection{INITIAL QUBIT-CAT STATE}
The results in the previous subsections are the case that the Wigner function of the reduced state of the oscillator is non-negative.
We here consider the cat state of the oscillator, where
its Wigner function has negative values and the negativity volume of the Wigner function is non-zero.
We investigate this model to verify
that the inequality \eqref{wig:criterion} can be used as an
entanglement criterion of a hybrid system.

The initial density matrix of the total system is
\begin{align}
	\label{q-cat:rho}
	\rho^{cat}(0)
	&=\ket{\phi^{cat}(0)}\bra{\phi^{cat}(0)},\qquad
	\ket{\phi^{cat}(0)}
	=\frac{1}{\sqrt{2N}}(\ket{0}+\ket{1})\otimes(\ket{\gamma}+\ket{-\gamma}),
\end{align}
where $N=2+2e^{-2|\gamma|^{2}}$.
The evolved state by unitary operator \eqref{hy:unitary} is given by
\begin{align}
	\label{q-cat:evo}
	\ket{\phi^{cat}(t)}
	&=\frac{1}{\sqrt{2N}}
	\left\{
	\ket{0}(e^{-(\alpha_{t}^{*}\gamma^{*}-\alpha_{t}\gamma)/2}\ket{\alpha_{t}+\gamma_{t}}+e^{(\alpha_{t}^{*}\gamma^{*}-\alpha_{t}\gamma)/2}\ket{\alpha_{t}-\gamma_{t}})\right.\notag\\
	&\left.\qquad\quad+
	\ket{1}(e^{(\alpha_{t}^{*}\gamma^{*}-\alpha_{t}\gamma)/2}\ket{-\alpha_{t}+\gamma_{t}}+e^{-(\alpha_{t}^{*}\gamma^{*}-\alpha_{t}\gamma)/2}\ket{-\alpha_{t}-\gamma_{t}})
	\right\},
\end{align}
and the generalized Wigner function is
\begin{align}
	\label{q-cat:W}
	W[\rho^{cat}(t)]
	&=\frac{1}{N\pi}
	\left[
	\frac{1-\sqrt{3}\cos\theta}{2}
	\left(
	e^{-2|\beta-\alpha_{t}-\gamma_{t}|^{2}}+e^{-2|\beta-\alpha_{t}+\gamma_{t}|^{2}}+2e^{-2|\beta-\alpha_{t}|^{2}}\cos[4\text{Im}[\alpha_{t}^{*}\gamma^{*}+\gamma_{t}^{*}\beta]]
	\right)\right.\notag\\
	&\left.\qquad\quad~
	+\frac{1+\sqrt{3}\cos\theta}{2}
	\left(
	e^{-2|\beta+\alpha_{t}-\gamma_{t}|^{2}}+e^{-2|\beta+\alpha_{t}+\gamma_{t}|^{2}}+2e^{-2|\beta+\alpha_{t}|^{2}}\cos[4\text{Im}[\alpha_{t}^{*}\gamma^{*}-\gamma_{t}^{*}\beta]]
	\right)\right.\notag\\
	&\left.\qquad\quad
	+\sqrt{3}\sin\theta\cos[4\text{Im}[\alpha_{t}^{*}\beta+\phi]]
	\left(
	e^{-2|\beta-\gamma_{t}|^{2}}+e^{-2|\beta+\gamma_{t}|^{2}}+2e^{-2|\beta|^{2}}\cos[4\text{Im}[\gamma_{t}^{*}\beta]]
	\right)
	\right].
\end{align}
Tracing over the qubit system, the reduced density matrix of the oscillator is derived by
\begin{align}
	\label{q-cat:rhored}
	\rho^{cat}_{b}(t)
	&=\frac{1}{2}(\rho^{cat}_{b1}(t)+\rho^{cat}_{b2}(t)),\\
	\rho^{cat}_{b1}(t)
	&=\frac{1}{N}
	\left(
	\ket{\alpha_{t}+\gamma_{t}}\bra{\alpha_{t}+\gamma_{t}}
	+e^{-(\alpha_{t}^{*}\gamma^{*}-\alpha_{t}\gamma)}\ket{\alpha_{t}+\gamma_{t}}\bra{\alpha_{t}-\gamma_{t}}\right.\notag\\
	&\left.\qquad~
	+e^{\alpha_{t}^{*}\gamma^{*}-\alpha_{t}\gamma}\ket{\alpha_{t}-\gamma_{t}}\bra{\alpha_{t}+\gamma_{t}}
	+\ket{\alpha_{t}-\gamma_{t}}\bra{\alpha_{t}-\gamma_{t}}
	\right),\\
	\rho^{cat}_{b2}(t)
	&=\frac{1}{N}
	\left(
	\ket{-\alpha_{t}+\gamma_{t}}\bra{-\alpha_{t}+\gamma_{t}}
	+e^{\alpha_{t}^{*}\gamma^{*}-\alpha_{t}\gamma}\ket{-\alpha_{t}+\gamma_{t}}\bra{-\alpha_{t}-\gamma_{t}}\right.\notag\\
	&\left.\qquad~
	+e^{-(\alpha_{t}^{*}\gamma^{*}-\alpha_{t}\gamma)}\ket{-\alpha_{t}-\gamma_{t}}\bra{-\alpha_{t}+\gamma_{t}}
	+\ket{-\alpha_{t}-\gamma_{t}}\bra{-\alpha_{t}-\gamma_{t}}
	\right),
\end{align}
and the Wigner function is
\begin{align}
	\label{q-cat:Wred}
	W_{b}[\rho^{cat}_{b1}(t)]
	&=\frac{2}{N\pi}
	\left(
	e^{-2|\beta-\alpha_{t}-\gamma_{t}|^{2}}
	+e^{-2|\beta-\alpha_{t}+\gamma_{t}|^{2}}
	+2e^{-2|\beta-\alpha_{t}|^{2}}\cos[4\text{Im}[\alpha_{t}^{*}\gamma^{*}+\gamma_{t}^{*}\beta]]\right),\\
	W_{b}[\rho^{cat}_{b2}(t)]
	&=\frac{2}{N\pi}
	\left(
	e^{-2|\beta+\alpha_{t}-\gamma_{t}|^{2}}
	+e^{-2|\beta+\alpha_{t}+\gamma_{t}|^{2}}
	+2e^{-2|\beta+\alpha_{t}|^{2}}\cos[4\text{Im}[\alpha_{t}^{*}\gamma^{*}-\gamma_{t}^{*}\beta]]
	\right).
\end{align}
Here, the Wigner function of each reduced state has negative regions because the reduced state is the non-Gaussian state.
Hence, the inequality for testing the entanglement is
\begin{align}
	\label{q-cat:criterion}
	\mathcal{V}[\rho^{cat}(t)]
	&\le
	\frac{2}{\sqrt{3}}\sum_{i=1}^{2}\frac{1}{2}\mathcal{V}[\rho^{cat}_{bi}(t)]+\frac{1}{\sqrt{3}}-\frac{1}{2}.
\end{align}
Figure \ref{fig:q-cat} shows the negativity volume (blue curve) and critical value (red dashed curve) for $\lambda=10^{-1}$ and $\gamma=1$.
The critical value (red dashed curve) is a function of time, but it is almost constant for the parameters.
The negativity volume is larger than the critical value except for 
$\omega t/\pi=2n \, (n=0,1,2,\dots)$.
Because the total system is in a pure state, the entanglement should be generated.
The result indicates that the entanglement generation due to the gravitational interaction is captured by the negativity volume of the generalized Wigner function.	

\begin{figure}[b]
	\includegraphics[width=8.4cm]{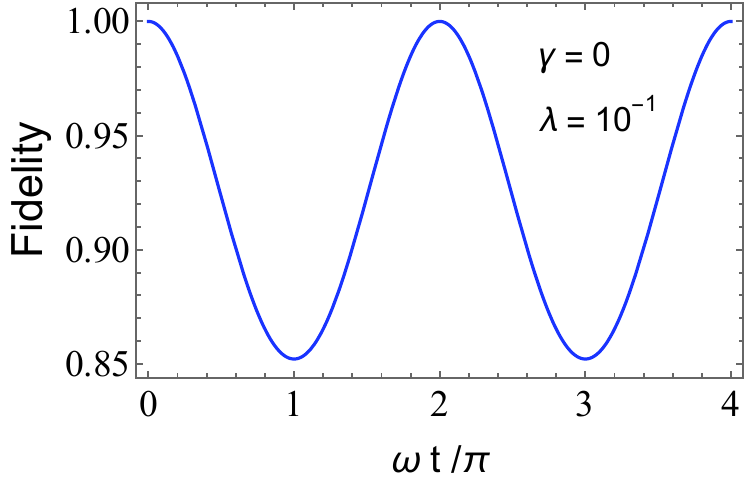}
\hspace{0.5cm}
 \includegraphics[width=8.6cm]{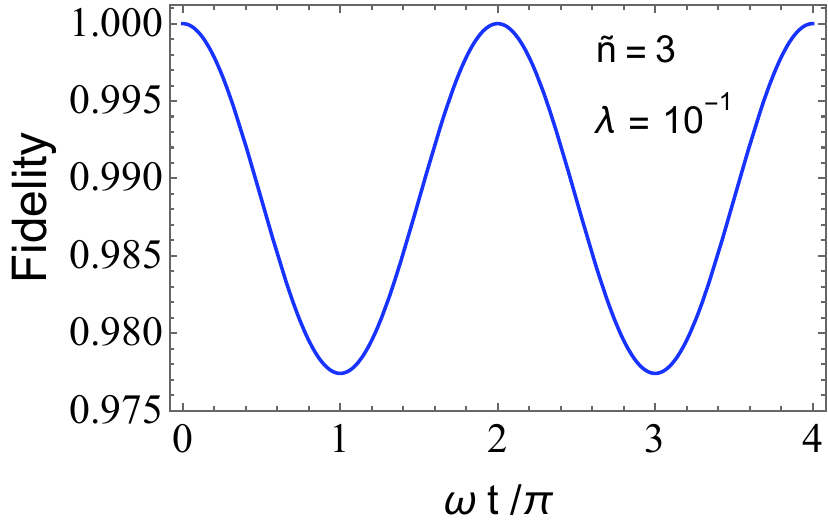}\\
 \vspace{5mm}
	\includegraphics[width=8.6cm]{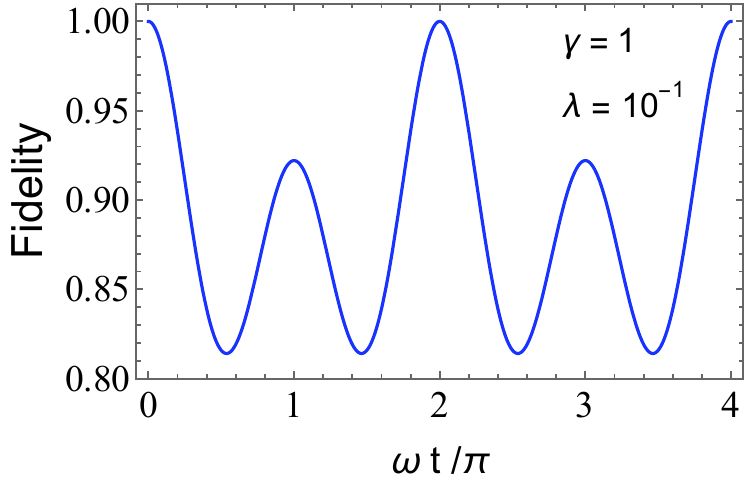}
	\caption{
	The behaviors of the fidelity for the initial qubit-coherent state (upper left), qubit-thermal state (upper right), and qubit-cat state (bottom), adopting the same parameters as Figs. \ref{fig:q-coh}, 
    \ref{fig:q-th}, and \ref{fig:q-cat}, respectively.
	The state is entangled if and only if the fidelity is not equal to 1.}
	\label{fig:fidelity}
\end{figure}
\section{ENTANGLEMENT FIDELITY}
In this section, we discuss how the evaluation of the entanglement by the Wigner function is consistent with that by the fidelity proposed in Ref.~\cite{Roszak15,Roszak18,Strzalka20}.
We here rewrite the Hamiltonian \eqref{hy:Hamiltonian} as
\begin{align}
	\hat{H}
	&=\omega\hat{a}^{\dagger}\hat{a}+\ket{0}\bra{0}\otimes g(\hat{a}+\hat{a}^{\dagger})+\ket{1}\bra{1}\otimes(-g(\hat{a}+\hat{a}^{\dagger})).
\end{align}
According to the entanglement criterion Ref. \cite{Roszak15,Roszak18,Strzalka20}, the state is separable if and only if the equality
\begin{align}
	\label{fid:criterion}
	[\rho_{b}(0),\hat{\omega}(t)]
	&=0
\end{align}
is satisfied, where the operator $\hat{\omega}(t)$ is given by
\begin{align}
	\label{fid:omega}
	\hat{\omega}(t)
	&=e^{i(\omega\hat{a}^{\dagger}\hat{a}+g(\hat{a}+\hat{a}^{\dagger}))t}
	e^{-i(\omega\hat{a}^{\dagger}\hat{a}-g(\hat{a}+\hat{a}^{\dagger}))t},
\end{align}
and we define the initial state as $\rho(0)=\rho_{q}^{p}(0)\otimes\rho_{b}(0)$.
This criterion is applicable to a initial product state of the pure qubit system and any other system, and is applicable to all the systems analyzed in the previous section.
Eq.~\eqref{fid:criterion} is rewritten as
\begin{align}
	\label{fid:criterion2}
	e^{-i(\omega\hat{a}^{\dagger}\hat{a}+g(\hat{a}+\hat{a}^{\dagger}))t}\rho_{b}(0)e^{i(\omega\hat{a}^{\dagger}\hat{a}+g(\hat{a}+\hat{a}^{\dagger}))t}
	&=e^{-i(\omega\hat{a}^{\dagger}\hat{a}-g(\hat{a}+\hat{a}^{\dagger}))t}\rho_{b}(0)e^{i(\omega\hat{a}^{\dagger}\hat{a}-g(\hat{a}+\hat{a}^{\dagger}))t}.
\end{align}
and this equation holds when the following operators $\sigma_{0}$ and $\sigma_{1}$ are equal:
\begin{align}
	\label{fid:sigma0}
	\sigma_{0}
	&=e^{-i\omega\hat{a}^{\dagger}\hat{a}t}e^{-i(\alpha_{t}^{*}\hat{a}^{\dagger}-\alpha_{t}\hat{a})}
	\rho_{b}(0)
	e^{i(\alpha_{t}^{*}\hat{a}^{\dagger}-\alpha_{t}\hat{a})}e^{i\omega\hat{a}^{\dagger}\hat{a}t},\\
	\label{fid:sigma1}
	\sigma_{1}
	&=e^{-i\omega\hat{a}^{\dagger}\hat{a}t}e^{i(\alpha_{t}^{*}\hat{a}^{\dagger}-\alpha_{t}\hat{a})}
	\rho_{b}(0)
	e^{-i(\alpha_{t}^{*}\hat{a}^{\dagger}-\alpha_{t}\hat{a})}e^{i\omega\hat{a}^{\dagger}\hat{a}t}.
\end{align}
To evaluate the identification of the operators $\sigma_{0}$ and $\sigma_{1}$, we introduce the fidelity as
\begin{align}
	\label{fidelity}
	F(\sigma_{0},\sigma_{1})
	&=\left(
	\text{Tr}\left[\sqrt{\sqrt{\sigma_{1}}\sigma_{0}\sqrt{\sigma_{1}}}\right]
	\right)^{2},
\end{align}
and the state is entangled if and only if Eq.~\eqref{fid:criterion2} is not satisfied, namely, $F\ne1$.

For the initial qubit-coherent state \eqref{q-coh:ini} in Sec.~III-A, we derive the fidelity as
\begin{align}
	\label{fid:q-coh}
	F(\sigma_{0},\sigma_{1})
	&=e^{-2|\alpha_{t}|^{2}},
\end{align}
which is not dependent on $\gamma$.
Similarly, the fidelity for the initial qubit-thermal state \eqref{q-th:ini} is obtained \cite{Marian} as
\begin{align}
	\label{fid:q-th}
	F(\sigma_{0}^{th},\sigma_{1}^{th})
	&=e^{-\frac{2}{2\bar{n}+1}|\alpha_{t}|^{2}},
\end{align}
and the fidelity for the initial qubit-cat state is
\begin{align}
	\label{fid:q-cat}
	F(\sigma_{0}^{cat}(t),\sigma_{1}^{cat}(t))
	&=\frac{1}{N}
	\left|
	e^{-2|\alpha_{t}-\gamma_{t}|^{2}}+e^{-2|\alpha_{t}+\gamma_{t}|^{2}}+2e^{-2|\alpha_{t}|^{2}}\cos\left(4\text{Im}[\alpha_{t}^{*}\gamma^{*}]\right)
	\right|,
\end{align}
where $N=2+2e^{-2|\gamma|^{2}}$.

Each panel of Fig.~\ref{fig:fidelity} plots the behavior of the fidelity \eqref{fid:q-coh}, \eqref{fid:q-th}, and \eqref{fid:q-cat}, as a function of time $\omega t/\pi$.
We find that the region where $F(\sigma_{0},\sigma_{1})\ne1$ is the same as the region where the negativity volume is larger than the critical value in Figs.~\ref{fig:q-coh},~\ref{fig:q-th},~\ref{fig:q-cat}.
Moreover, the fidelity for the qubit-thermal state is closer to $1$ than that of the qubit-coherent state, which corresponds to the fact that the negativity volume of the qubit-thermal state is smaller than that of the initial qubit-coherent state. Thus the comparison of the conditions for quantum entanglement
demonstrated the consistency between the method with the negativity volume of the generalized Wigner function and the method with fidelity, both of which properly detect the quantum entanglement in the hybrid system.
In all these models, the signature of entanglement is completely captured by the negativity volume of the generalized Wigner function.

\section{SUMMARY AND CONCLUSION}
We examined the quantum nature of gravity in terms of the generalized Wigner function in a hybrid system consisting of a two-level particle and an oscillator.
The two-level particle means a spatially localized superposed state at two different points and the total system evolves into nonclassical entangled states through gravitational interaction.
We discussed the entanglement due to gravity by using the quantum entanglement criterion, which states that an increase in the negativity volume of the generalized Wigner function corresponds to quantum entanglement, as proposed in Ref. \cite{Arkhipov18}.
Although this criterion is not a necessary condition, we demonstrated that the entanglement is adequately captured by comparing it with the condition using the fidelity.
Our results suggest that the criterion with the generalized Wigner function is valid for detecting the gravity-induced entanglement, though the increase in the negativity volume is small in a realistic situation because gravity is weak.

\acknowledgments

D. Miki thanks Prof. Kae Nemoto for her hospitality during his stay at OIST and for useful discussions on the topic of the present paper. 
D. M. is supported by JSPS KAKENHI, Grant No. 22J21267, K.Y. is supported by JSPS KAKENHI,
Grant No. JP22H05263 and No. JP23H01175.

\appendix
\section{ENTANGLEMENT CRITERION BASED ON THE GENERALIZED WIGNER FUNCTION}
We explain the entanglement criterion proposed in Ref. \cite{Arkhipov18}.
Here, we consider the separable state of a hybrid system $\rho^{\text{sep}}=\sum_{i}p_{i}\rho_{q}^{i}\otimes\rho_{b}^{i}$, where $p_{i}$ is positive values satisfying $\sum_{i}p_{i}=1$.
The generalized Wigner function of the separable system is
\begin{align}
\label{ap:wig}
W[\rho^{\text{sep}}]
&=\sum_{i}p_{i}\text{tr}[\rho_{q}^{i}\hat{\Delta_{q}}]\text{tr}[\rho_{b}^{i}\hat{\Delta_{b}}]
=\sum_{i}p_{i}W_{q}[\rho_{q}^{i}]W_{b}[\rho_{b}^{i}],
\end{align}
and the negativity volume is
\begin{align}
\label{ap:nv}
\mathcal{V}[\rho^{\text{sep}}]
&=\frac{1}{2}\left(\int d\phi d\theta d^{2}\beta\frac{\sin\theta}{2\pi}|W[\rho^{\text{sep}}]|-1\right)\notag\\
&\le\frac{1}{2}\left(\sum_{i}p_{i}\int d\phi d\theta d^{2}\beta\frac{\sin\theta}{2\pi}|W_{q}[\rho_{q}^{i}]W_{b}[\rho_{b}^{i}]|-1\right)\notag\\
&=\frac{1}{2}\sum_{i}p_{i}
\left\{
\left(\int d\phi d\theta\frac{\sin\theta}{2\pi}|W_{q}[\rho_{q}^{i}]|-1\right)
\left(\int d\beta|W_{b}[\rho_{b}^{i}]|-1\right)\right.\notag\\
&\left.\quad
+\left(\int d\phi d\theta\frac{\sin\theta}{2\pi}|W_{q}[\rho_{q}^{i}]|-1\right)
+\left(\int d\beta|W_{b}[\rho_{b}^{i}]|-1\right)
\right\}\notag\\
&=\sum_{i}p_{i}\left(2\mathcal{V}_{q}[\rho_{q}^{i}]\mathcal{V}_{b}[\rho_{b}^{i}]+\mathcal{V}_{q}[\rho_{q}^{i}]+\mathcal{V}_{b}[\rho_{b}^{i}]\right).
\end{align}
Since the negativity volume of the Wigner function for the qubit system is upper bounded by the pure state, $1/\sqrt{3}-1/2$, then the negativity volume of the separable state of the hybrid system satisfies the following inequality:
\begin{align}
\label{ap:criterion}
\mathcal{V}[\rho^{\text{sep}}]
&\le \frac{2}{\sqrt{3}}\sum_{i}p_{i}\mathcal{V}[\rho_{b}^{i}]+\frac{1}{\sqrt{3}}-\frac{1}{2}.
\end{align}
Hence, if the inequality is violated, it means the nonclassicality of the hybrid system.
We note that this criterion is the only sufficient criterion for the entanglement generation and there are cases where the inequality \eqref{ap:criterion} is satisfied even though the state is quantum entangled, as shown in Ref. \cite{Arkhipov18}.
Also, there is no restriction of the negativity volume for the reduced continuous variable system $\mathcal{V}_{b}[\rho_{b}^{i}]$.

\end{document}